\documentclass{article}

\usepackage[english]{babel}
\usepackage{graphicx}
\usepackage{subfigure}
\usepackage[bottom]{footmisc}
\usepackage{float}
\usepackage{dblfloatfix}

\makeatletter
\let\@fnsymbol\@arabic
\makeatother

\begin{document}

\title{Study of prototypes of LFoundry active and monolithic CMOS pixels sensors for the ATLAS detector}
\author{L. Vigani\footnote{University of Oxford} \and D. Bortoletto\footnotemark[1] \and L. Ambroz\footnotemark[1] \and R. Plackett\footnotemark[1] \and T. Hemperek\footnote{University of Bonn} \and P. Rymaszewski\footnotemark[2] \and T. Wang\footnotemark[2] \and H. Krueger\footnotemark[2] \and T. Hirono\footnotemark[2] \and I. Caicedo Sierra\footnotemark[2] \and N. Wermes\footnotemark[2] \and M. Barbero\footnote{CPPM - CNRS/IN2P3 / Aix-Marseille Universit\'e} \and S. Bhat\footnotemark[3] \and P. Breugnon\footnote{Centre National de la Recherche Scientifique} \and Z. Chen\footnotemark[3] \and S. Godiot\footnotemark[4] \and P. Pangaud\footnotemark[3] \and A. Rozanov\footnotemark[3]
}
\date{}

\maketitle

\begin{abstract}
High Energy Particle Physics experiments at the LHC use hybrid silicon detectors, in both pixel and strip geometry, for their inner trackers. These detectors have proven to be very reliable and performant. Nevertheless, there is great interest in the development of depleted CMOS silicon detectors, which could achieve similar performances at lower cost of production and complexity. We present recent developments of this technology in the framework of the ATLAS CMOS demonstrator project. In particular, studies of two active sensors from LFoundry, CCPD$\_$LF and LFCPIX, and the first fully monolithic prototype MONOPIX will be shown.
\end{abstract}

\newpage

\section{The ATLAS upgrade}

\subsection{The ATLAS experiment}
ATLAS (A Large Toroidal LHC ApparatuS) is a large multi-purpose High Energy Particle Physics experiment at the Large Hadron Collider (LHC) at CERN. It consists of several detector elements, nested around the interaction point, measuring the properties of the particles produced in the LHC collisions. The main goals of the ATLAS physics program are precision studies of the Standard Model Physics, including the recently found Higgs Boson, and searches for physics beyond the standard model ,including Supersymmetry.\\

\subsection{Towards HL-LHC}
\begin{figure}[H]
\centering
\includegraphics[scale=0.15]{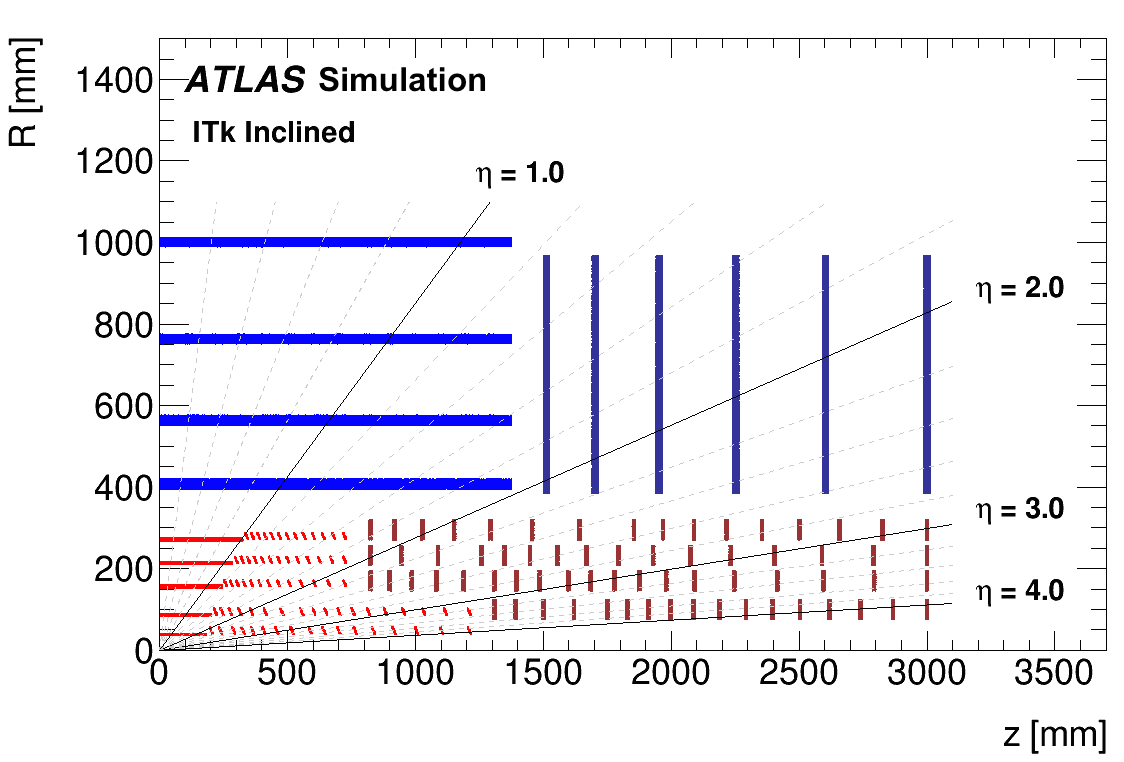}
\caption{(Left) Sketch of the ATLAS experiment. (Right) One of the possible layouts of the ItK system. Pixel modules are in red, while strip modules in blue.}
\label{fig:itk_layout}
\end{figure}

The Large Hadron Collider, is currently taking data at $13 TeV$ and at a luminosity of $10^{34} cm^{-2} s^{-1}$. It is expected that the center of mass energy of the machine will increase to $14 Tev$ for run 3. Then, the machine is scheduled to undergo an upgrade, denoted as the High Luminosity LHC or HL-LHC, that will increase the instantaneous luminosity to about $7 \cdot 10^{34} cm^{-2} s^{-1}$. In this phase, the bunch crossing time will remain $25 ns$ , while the number of proton collision per bunch crossing will be increased. This luminosity increase will set new requirements to the detectors in terms of speed, data acquisition rate and radiation resistance. In particular, the tracking system, the innermost part of the experiment, will be replaced with an all Silicon tracker, the ITK. This will be divided in two parts, the inner tracker, made of Silicon pixel detectors, and the outer tracker, made of Silicon strip detectors.\\
The devices presented in this paper are potential candidates for the 5th (last) layer of the pixel system, at about 30 $cm$ from the interaction vertex (see Figure ~\ref{fig:itk_layout}). At that distance the fluence and the Total Ionization Dose (TID) are expected to be $1.5 \cdot 10^{15} n_{eq}/cm^2$ and $80 Mrad$ respectively.

\section{Introduction to CMOS detectors in ATLAS}

\subsection{Silicon Hybrid detectors}

\begin{figure}
\centering
\includegraphics[scale=0.12]{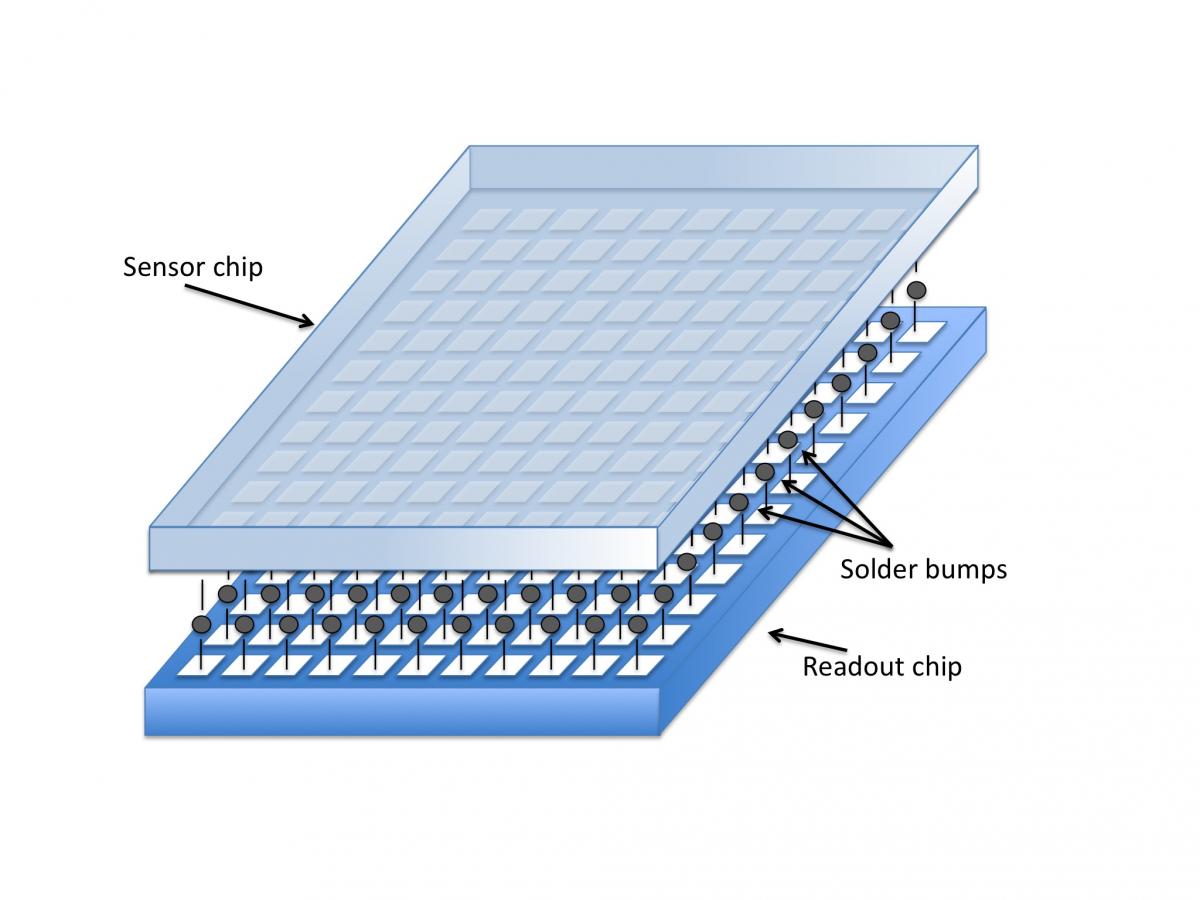}
\caption{Hybrid pixel detector sketch. (Right) Lateral layout of the standard CMOS process.}
\label{fig:hybrids}
\end{figure}

Most current HEP experiments at the LHC have a pixel system based on Silicon Hybrid technologies. These are produced by coupling a high resistivity Silicon diode with a front end read-out chip. The connection is guaranteed by arrays of conductive bumps, one per pixel, generally made with Indium. Picture ~\ref{fig:hybrids} shows a sketch of this process.\\
These detectors have achieved excellent performances providing good signal response within $25 ns$. They are also radiation hard up to $5 \cdot 10^{15} n_{eq}/cm^2$. Nevertheless, this technology, although very well understood, remains very difficult and expensive to produce, since the bump bonding process is slow, complicated and with high risk of failure. For this reason, there is a lot of interest for novel detectors called depleted Monolithic Active Pixel Sensors (DMAPS).

\subsection{CMOS detectors: from standard applications to HEP}

\begin{figure}
\centering
\subfigure[]{\includegraphics[scale=0.25]{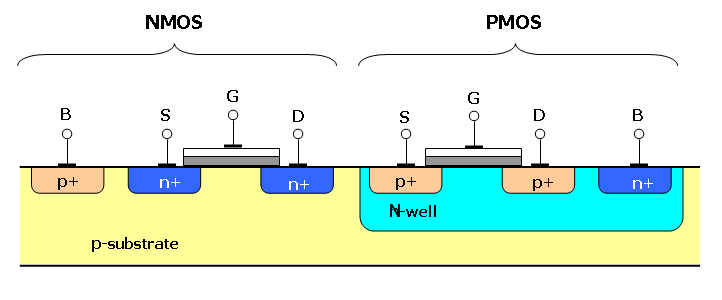}\label{fig:standard_cmoss}}
\subfigure[]{\includegraphics[scale=0.4]{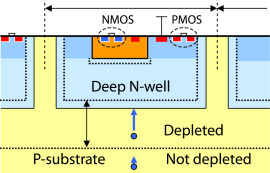}\label{fig:hv_cmoss}}
\caption{(Left) Standard CMOS sensor layout (from side). (Right) Layout of the CMOS modified process for high bias voltage.}
\end{figure}

The CMOS technology provides a different concept of detector, in which the read-out circuitry is inserted inside the Silicon sensor itself (for this reason they are also called MAPS). This allows an easy production at lower cost, with less material. These sensors has been used in a variety of applications from industry to Medical Physics (see Figure ~\ref{fig:standard_cmoss}).\\
Standard MAPS detectors are not suitable for LHC and HL-LHC operation, since in these detectors the charge collection happens trough thermal diffusion, which leads to a charge collection time that is too long for these challenging environments. For this reason, some improvements must be implemented in the technology. Specifically, the resistivity of the substrate and the bias voltage must be increased, since the depletion region and therefore the charge collected is proportional to $\sqrt{\rho \cdot V}$. The read-out circuitry should be nested inside a deep N-well, providing isolation to operate at high voltages. This way, the depletion region is formed between the substrate and the deep N-well itself \cite{peric}. A sketch of these add-ons can be seen in Figure ~\ref{fig:hv_cmoss}.

\subsection{The ATLAS CMOS pixel collaboration}

The ATLAS demonstrator program aims at proving the feasibility of the DMAPS technology for the ITk. Different foundries and strategies have been investigated. One of the main feature to study is the fill factor, i.e. the ratio between the surface of the deep N-well, shown in Figure ~\ref{fig:hv_cmoss}, and the total surface of the pixel. In particular, with a large fill factor the read-out is inside the collection well, while with the small fill factor it is outside. The first option can achieve higher bias voltages and  signals but is affected by higher noise. The second option is optimized to minimize the capacitance and therefore the noise and the power dissipated.\\
The small fill factor approach has been developed in the design and production of prototypes with TowerJazz foundry, while the large fill factor has been implemented with LFoundry and \textit{ams}. Herein we describe the current status of the studies performed to evaluate the LFoundry  process.

\section{LFoundry prototypes}

In the past three years, three prototypes have been designed and fabricated in the LFoundry technology: CCPD$\_$LF, LF-CPIX and LF-MONOPIX.\\

\subsection{Active CMOS devices}

\begin{figure}
\centering
\includegraphics[scale=0.25]{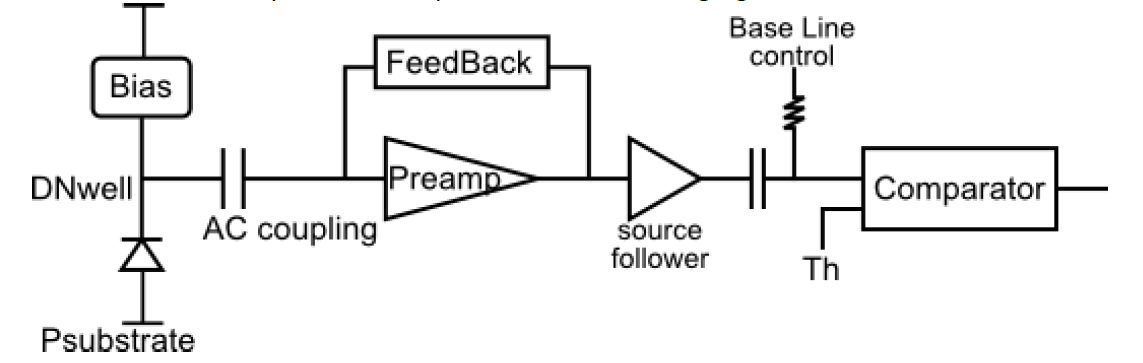}
\caption{Read-out schematics for the active CMOS sensors.}
\label{fig:active_ro}
\end{figure}

CCPD$\_$LF and LF-CPIX have a very similar design: in both cases amplifiers,  discriminators and other signal processing are implemented insides a deep N-well. For this reason they are referred to as active CMOS devices. In addition, both have different kinds of pre-amplifier: PMOS, NMOS and CMOS (the last only for LF-CPIX)\cite{toko}. The basic read-out scheme is sketched in Figure ~\ref{fig:active_ro}.

\begin{figure}
\centering
\subfigure[]{\includegraphics[scale=0.25]{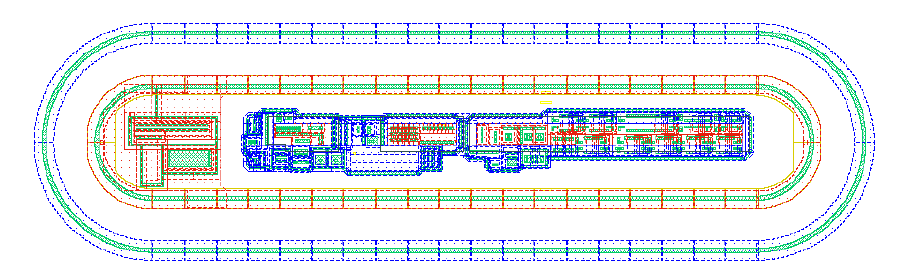}\label{fig:ccpd_layout}}
\subfigure[]{\includegraphics[scale=0.3]{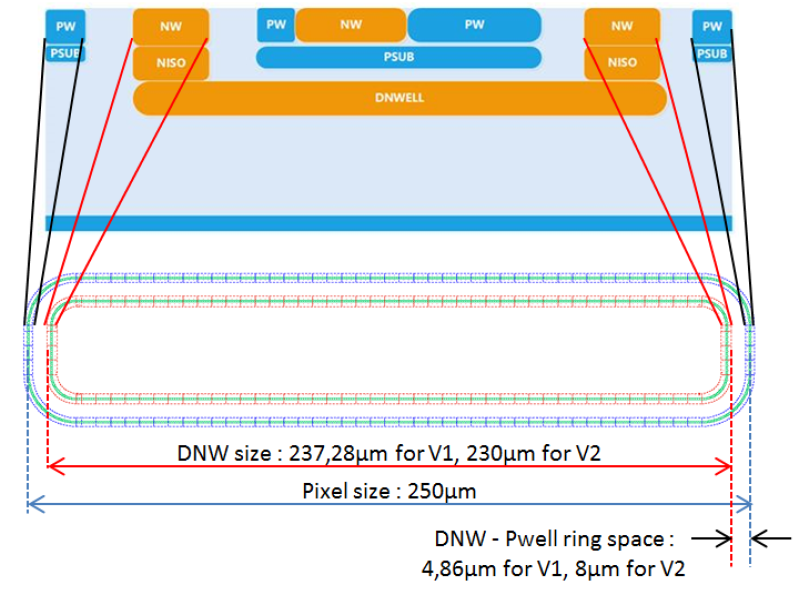}\label{fig:cpix_layout}}
\caption{(a) Front layout of CCPD$\_$LF. (b) Lateral and front layout of LF-CPIX.}
\end{figure}

The main difference between the two devices is the pixel size, which is $33 \times 125 \mu m^2$ for CCPD$\_$LF and $50 \times 250 \mu m^2$ for LF-CPIX, the latter being the standard ATLAS pixel size (same as FE-I4). The layout of the two active pixel prototypes can be seen in Figures ~\ref{fig:ccpd_layout} and ~\ref{fig:cpix_layout}. Looking at the front layout, one can easily notice the large fill factor as the deep N-well occupies most of the pixel surface. Figure ~\ref{fig:cpix_layout}, in particular, show also the lateral layout, where the multiple nested wells can be seen. Both devices can be either capacitively glued to an FE-I4 chip or read-out through a standalone board.

\subsection{Fully Monolithic device}

\begin{figure}
\centering
\includegraphics[scale=0.25]{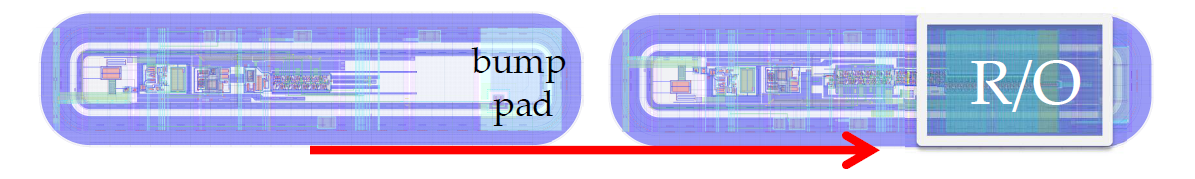}
\caption{Sketch of the design transition from LF-CPIX to LF-MONOPIX.}
\label{fig:monopix_sketch}
\end{figure}

The last prototype, LF-MONOPIX, is a fully monolithic device, i.e. it implements the whole read-out inside the pixel itself. As shown in Figure ~\ref{fig:monopix_sketch}, the digital part of the read-out, which follows the FE-I3 architecture, was added on the right-hand side of the analogue pixel based on LF-CPIX. LF-MONOPIX delivers a serialized output with a 40MHz  clock ($25 ns$  resolution) and no read-out chip is needed. Encoded in the output bit-stream are the time-stamps of the leading edge and the trailing edge of each hit, from which one can calculate the Time over Threshold (ToT) and hence the charge collected.\\
Several LF-MONOPIX devices are under study and results will be published soon.

\section{Characterization and results}

CCPD$\_$LF, LF-CPIX and LF-MONOPIX have been extensively characterized within the collaboration with different tests and irradiation doses and sources. Some results are shown in this section, while others are currently being collected and processed.

\subsection{CCPD$\_$LF performances before irradiation}

\begin{figure}[H]
\centering
\subfigure[]{\includegraphics[scale=0.22]{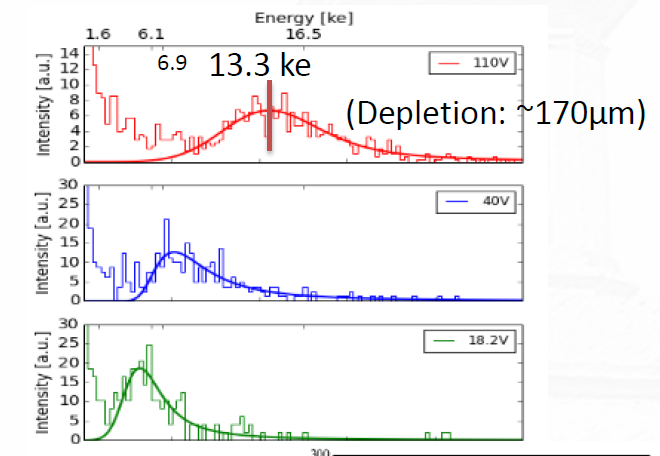}\label{fig:ccpd_landaus}}
\subfigure[]{\includegraphics[scale=0.3]{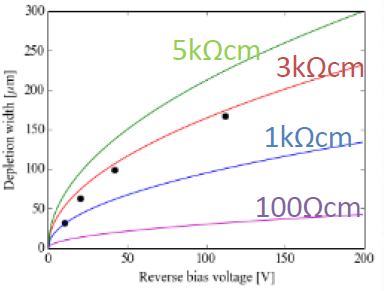}\label{fig:ccpd_rho}}
\caption{(a) Spectra obtained with CCPD$\_$LF at different biase voltages. (b) MPV as a function of bias and square root fit.}
\end{figure}

\begin{figure}[H]
\centering
\subfigure[]{\includegraphics[scale=0.13]{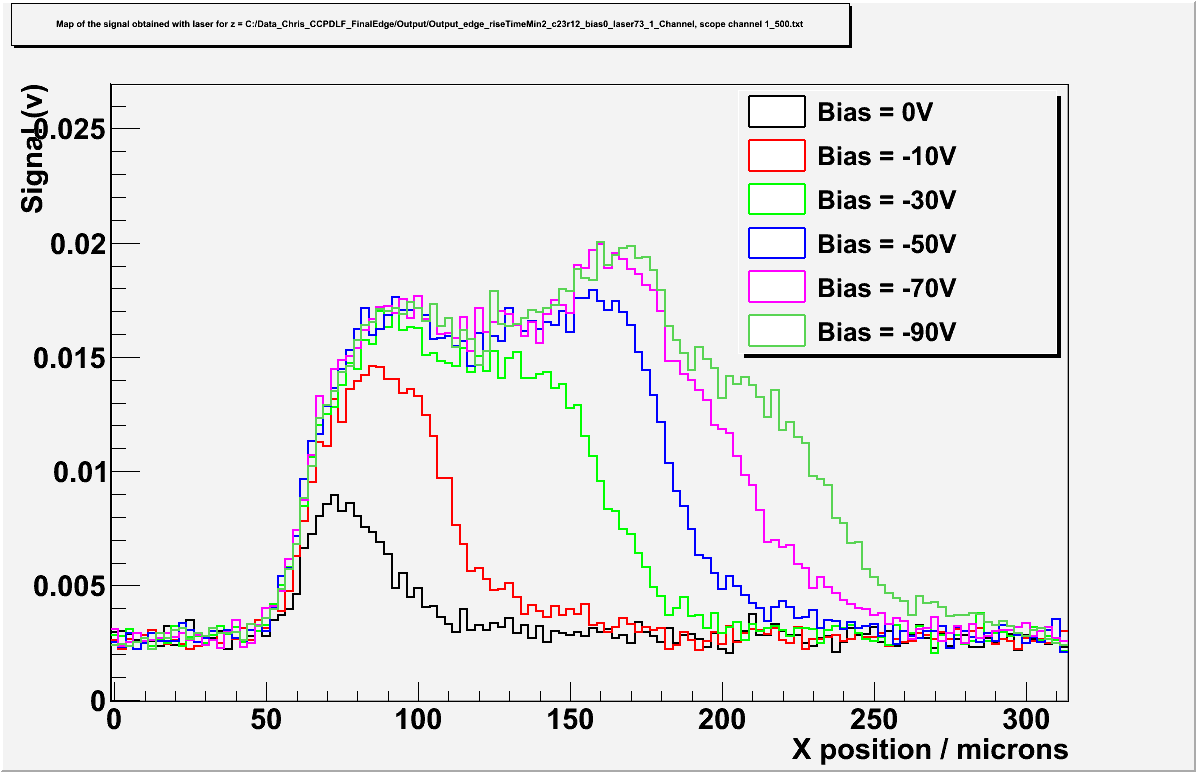}\label{fig:ccpd_etct}}
\subfigure[]{\includegraphics[scale=0.21]{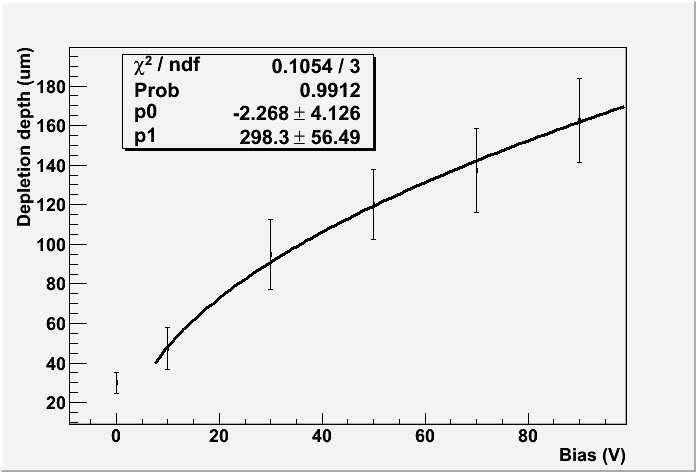}\label{fig:ccpd_etct_fit}}
\caption{(a) Edge-TCT 1D scans of a CCPD$\_$LF active pixel. (b) FWHM of the signal profile vs bias.}
\end{figure}

CCPD$\_$LF characterization with MIPs was performed at ELSA facility in Bonn with a $3.2 GeV$ electron beam. The collected spectra show a distribution compatible with a Landau (see Figure ~\ref{fig:ccpd_landaus}), and the dependence of the Maximum Probability Value (MPV) parameter on the bias is compatible with a resistivity of $3 k \Omega \cdot cm$ (see Figure ~\ref{fig:ccpd_rho}).\\
This result has been confirmed by investigating the properties of charge collection with edge-TCT, a technique that consists in shining very fast Infra-Red laser pulses on the edge of the sensor\cite{etct}. Once the laser is well focused, one can study the signal produced at different depth levels inside the bulk and have a direct measurement of the depletion depth. Figure ~\ref{fig:ccpd_etct} shows the charge profiles obtained with this method at different bias voltages, showing  that the depletion region and therefore the signal increases with bias voltage as expected. The depletion depth is then calculated as the FWHM of this profile, and its value is plotted as a function of the bias voltage in Figure ~\ref{fig:ccpd_etct_fit}. From an interpolation with a square root function it was possible to estimate the resistivity to be $3.3 \pm 0.5 k \Omega \cdot cm$, which confirms the result obtained with an electron beam.


\subsection{LF-CPIX performances before irradiation}

\begin{figure}[H]
\centering
\subfigure[]{\includegraphics[scale=0.13]{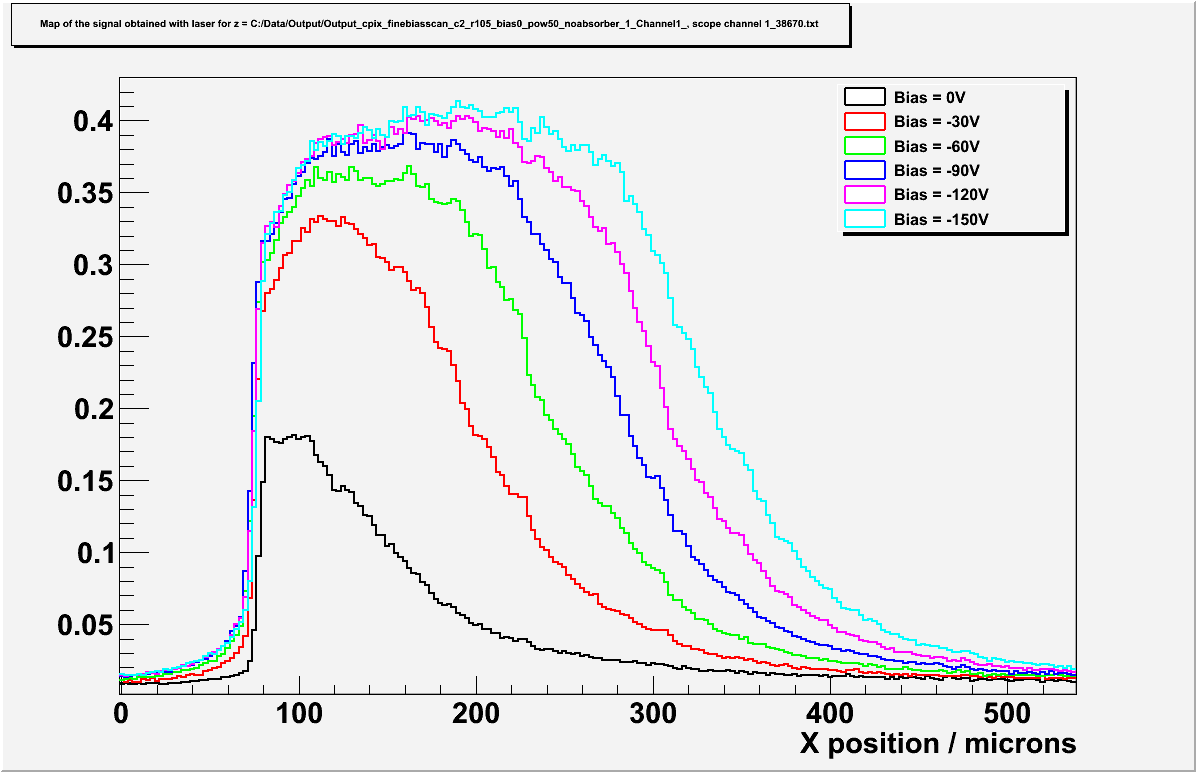}\label{fig:cpix_etct}}
\subfigure[]{\includegraphics[scale=0.21]{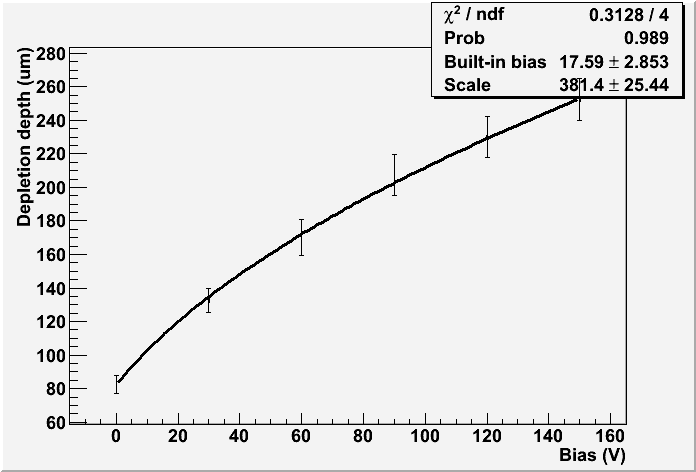}
\label{fig:cpix_etct_fit}}
\caption{(a) Edge-TCT 1D scans of an LF-CPIX active pixel. (b) FWHM of the signal profile vs bias.}
\end{figure}

In the same way as CCPD$\_$LF, the properties of charge collection for LF-CPIX have been investigated with edge-TCT. Figure ~\ref{fig:cpix_etct} shows the charge collection profile as a function of the laser depth for different bias voltages. From an interpolation with a square root function (Figure ~\ref{fig:cpix_etct_fit}) it was possible to estimate the resistivity to be $4.2 \pm 0.3 k \Omega \cdot cm$.\\


\subsection{CCPD$\_$LF performances after neutron irradiation}

\begin{figure}[H]
\centering
\includegraphics[scale=0.25]{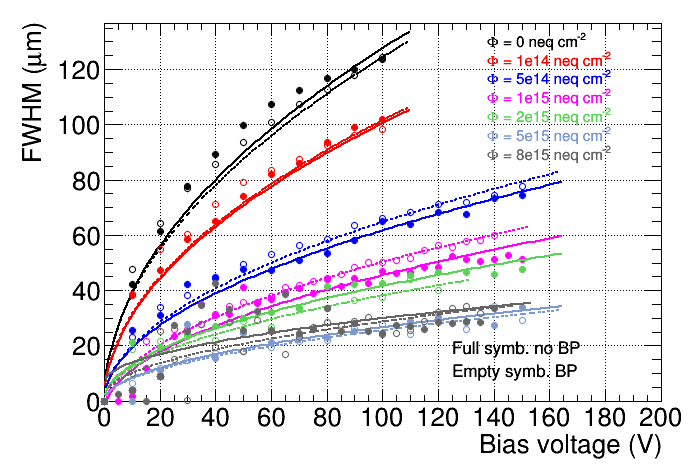}
\caption{Depletion depth measured with edge-TCT at different biases and neutron irradiation levels.}
\label{fig:ccpd_neutron}
\end{figure}

Some CCPD$\_$LF prototypes have been irradiated with neutrons in Ljubljana up to a fluence of $5 \cdot 10^{15} n_{eq} cm^{-2}$ and then tested with the edge-TCT to evaluate the reduction in the depletion region due to bulk damage. The results are shown in Figure ~\ref{fig:ccpd_neutron}. From this plot, it is clear that even after a fluence of $2 \cdot 10^{15} n_{eq} cm^{-2}$ a depletion depth of more than $50 \mu m$  can be obtained, surpassing the requirements for operation in the outer pixel layers.

\subsection{LF-CPIX performances after proton irradiation}

\begin{figure}[H]
\centering
\subfigure[]{\includegraphics[scale=0.13]{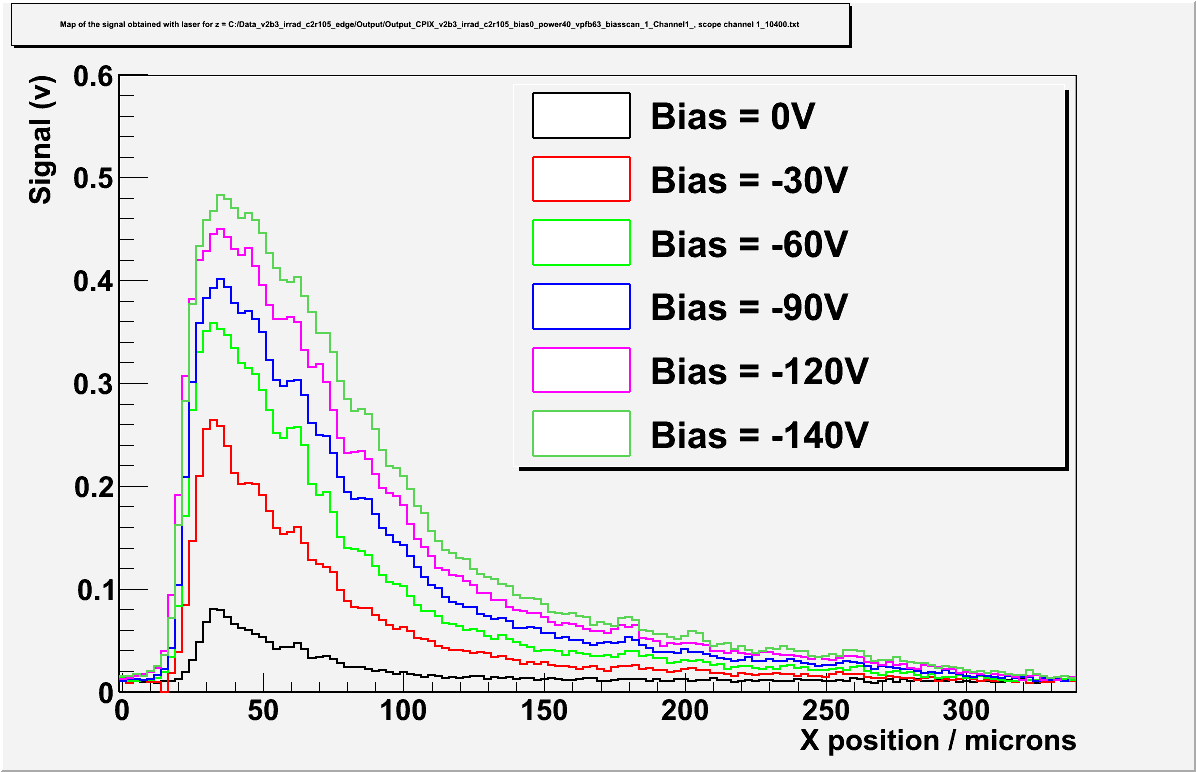}\label{fig:cpix_irrad_etct}}
\subfigure[]{\includegraphics[scale=0.21]{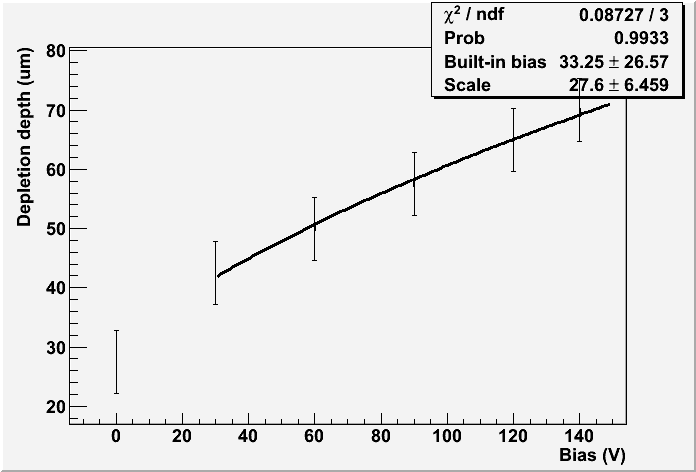}\label{fig:cpix_irrad_etct_fit}}
\caption{(a) Edge-TCT 1D scans of an LF-CPIX active pixel after proton irradiation. (b) FWHM of the signal profile vs bias.}
\end{figure}

\begin{figure}[H]
\centering
\includegraphics[scale=0.25]{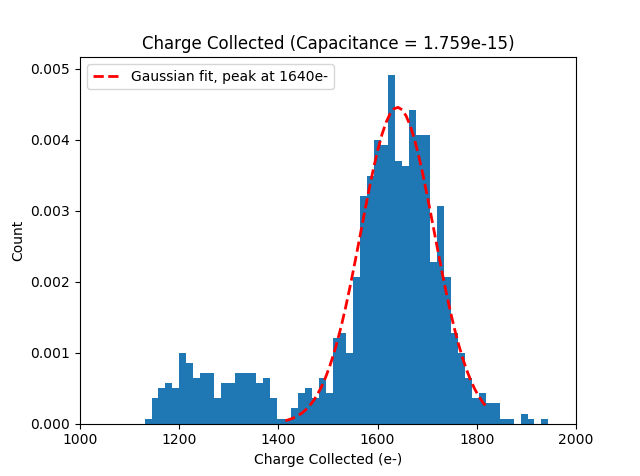}
\caption{$^{55}Fe$ spectrum obtained with proton irradiated LF-CPIX}
\label{fig:cpix_irrad_fe55}
\end{figure}

Some LF-CPIX samples have been irradiated with 27 $MeV$ protons to a fluence of $10^{15} n_{eq} cm^{-2}$ at the Birmingham Cyclotron\cite{Bham}. Afterwards, edge-TCT has been performed to check the effect of the irradiation on the charge collection. The results, shown in Figures ~\ref{fig:cpix_irrad_etct} and ~\ref{fig:cpix_irrad_etct_fit}, prove that the device can be depleted to a depth of more than 70 $\mu m$ after this dose.\\
In addition, an $X$-ray characterization with $^{55}Fe$ proves a good behaviour of the device (Figure ~\ref{fig:cpix_irrad_fe55}), as the expected peak at $1640 e^-$ has been observed.

\section{Conclusion}

The ATLAS upgrade is going to impose very high standards on the performance of its new tracking system. A new concept of Silicon detectors, based on the CMOS technology, is under investigation as a candidate for the fifth pixel layer. The new  technology brings great advantages and allow easy and low cost of production of detector modules. The ATLAS demonstrator program is studying different prototypes and foundries to prove the feasibility of this technology for the ATLAS upgrade.\\
One of the foundries that have been investigated is LFoundry, which provides sensors with high resistivity and high voltage additions. These add-ons ensure a sufficiently high signal. Three devices have been designed and produced with this foundry, two active CMOS sensors, CCPD$\_$LF and LF-CPIX, and a fully monolithic device, LF-MONOPIX. Many tests have been performed on these devices before and after irradiation with with neutrons, protons and $X$-rays\\
The tests performed so far indicate that these devices are fully functional at the needed high radiation doses. The next step for the collaboration is then to produce a full CMOS module to finally prove the feasibility of this technology for the ATLAS upgrade.

\section*{Acknowledgments}
\begin{figure}[H]
\centering
\includegraphics[width=\textwidth]{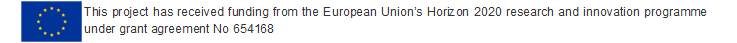}
\end{figure}

\end{document}